\documentclass{eas}
\usepackage{graphicx}
\newcommand{\ha}{H\ensuremath{\alpha}}
\newcommand{\hii}{H~{\sc ii}}
\newcommand{\hi}{H~{\sc i}}

\newcommand{\sii}{[S~{\sc ii}]}
\newcommand{\nii}{[N~{\sc ii}]}

\newcommand{\oii}{[O~{\sc ii}]}

\begin{document}

\title{Diffuse Ionized Gas in Spiral Galaxies and the Disk-Halo 
Interaction} 
\runningtitle{Reynolds \etal: Diffuse Ionized Gas in Galaxies}
\author{R. J. Reynolds}\address{Astronomy Department, University of 
Wisconsin, Madison WI 53706 USA}
\author{L. M. Haffner}\sameaddress{1}
\author{G. J. Madsen}\address{School of Physics, The University of 
Sydney, NSW 2006, Australia}
\author{K. Wood}\address{School of Physics and Astronomy, University of 
St. Andrews, Scotland}
\author{A.~S.~Hill}\sameaddress{1}

\begin{abstract}

Thick layers of warm, low density ionized hydrogen (i.e., the warm ionized 
medium or WIM) in spiral galaxies provide direct evidence for an 
interaction between the disk and halo.  The wide-spread ionization implies 
that a significant fraction of the Lyman continuum photons from O~stars, 
produced primarily in isolated star forming regions near the midplane and 
often surrounded by opaque clouds of neutral hydrogen, is somehow able to 
propagate large distances through the disk and into the halo. Moreover, 
even though O~stars are the source of the ionization, the temperature and 
ionization state of the WIM differ significantly from what is observed in 
the classical O~star \hii\ regions. Therefore, the existence of the WIM 
and observations of its properties provide information about the structure 
of the interstellar medium and the transport of energy away from the 
midplane as well as place significant constraints on models.

\end{abstract}

\maketitle

\section{Introduction}

Forty-five years ago, Hoyle \& Ellis (\cite{HE63}) proposed the existence 
of an extensive layer of warm (10$^4$~K), low density 
(10$^{-1}$~cm$^{-3}$)  ionized hydrogen surrounding the plane of our 
Galaxy and having a power requirement comparable to the ionizing 
luminosity of the Galaxy's O and B stars. Their conclusion was based upon 
their discovery of a free-free absorption signature in the observations of 
the Galactic synchrotron background at frequencies below 10 MHz. However, 
the idea that a significant fraction of the Lyman continuum photons from 
the Galaxy's O~stars could travel hundreds of parsecs throughout the disk 
conflicted with the traditional picture in which the interstellar neutral 
hydrogen confined the ionizing radiation to small volumes (i.e., 
``classical \hii\ regions'') near the hot stars.  Nevertheless, a few 
years later, the dispersion of radio signals from newly discovered pulsars 
(Hewish {\em et al.\/} \cite{HBP68}) plus the detection of faint optical 
emission lines from the diffuse interstellar medium (Reynolds 
\cite{Reynolds71}) firmly established warm ionized hydrogen as a major, 
wide-spread component of our Galaxy's interstellar medium.  But it was an 
additional two decades before deep \ha\ imaging with CCDs began to reveal 
similar warm plasmas permeating the disks and halos of other galaxies 
(Rand {\em et al.\/} \cite{RKH90}; Dettmar \cite{Dettmar90}), establishing 
the WIM as not just a peculiarity of the Milky Way, but a common property 
of galaxy disks.

\section{Basic Properties of the WIM}

Though originally detected by radio techniques, subsequent developments in 
high-throughput Fabry-Perot spectroscopy and CCD imaging have demonstrated 
that the primary source of information about the distribution, kinematics, 
and other physical properties of the WIM is through the detection and 
study of its faint emission lines at optical wavelengths. Not 
surprisingly, the most detailed studies have been made for the Milky Way, 
particularly for the region within about 3 kpc of the sun.  For example, 
the Wisconsin \ha\ Mapper (WHAM) has been used to map the distribution and 
kinematics of the H$^+$ over the sky through the hydrogen recombination 
line and to probe the temperature and ionization conditions through the 
detection of weaker forbidden lines of trace ions and atoms (e.g., Haffner 
{\em et al.\/} \cite{WHAMNSS}; Madsen {\em et al.\/} \cite{MRH06}).

The WHAM survey has revealed a complex morphology for the H$^+$, with 
filaments and blobs extending to high Galactic latitudes and superposed on 
a smoother \ha\ background that covers the sky.  Comparisons of emission 
measures ($\propto n^2$), derived from the \ha\ intensity, with pulsar 
dispersion measures ($\propto n$) indicate that the H$^+$ is clumped into 
regions having an average electron density $n_{e} = 0.03 - 0.08$ cm$^{-3}$ 
and filling a fraction $f \approx 0.4 - 0.2$ of the volume within a $2000 
- 3000$ pc thick layer about the Galactic midplane.  Half of the H$^+$ is 
located at heights $|z| > 700$ pc.  Within these ionized regions, the 
hydrogen is nearly fully ionized (H$^+$/H $> 0.9$) and has a temperature 
that is generally warmer than that of classical O~star \hii\ regions 
(see Sec. 4.2).  The WIM accounts for 90\% or more of the ionized 
hydrogen within the interstellar medium, and along lines of sight at 
high Galactic latitude, the column density of the \hii\ is observed to 
range from 20\% to 60\% that of the \hi. For details on how these 
parameters were derived, see the review by Haffner {\em et al.\/} 
(\cite{haffner09}) and references therein.

\section{O~Stars as the Source of the Ionization}

The low \ha\ surface brightness of High Velocity Clouds (HVCs) in the halo 
of the Galaxy, approximately 1/10 that associated with the WIM (e.g., 
Tufte {\em et al.\/} \cite{TRH98}), implies that the source of the 
ionization of the WIM must be within the disk. In the solar neighborhood, 
the H-ionization rate needed to sustain the WIM is about 1/7 that 
available from stellar Lyman continuum photons. In other galaxies, the 
fractions are found to scatter around a value closer to 1/2.  Only O~stars 
have this much ionizing power (e.g., Reynolds \cite{Reynolds90}).  
Studies of face-on galaxies in particular reveal that the total \ha\ 
luminosity of the diffuse ionized gas is on average approximately equal to 
that of the discrete, classical \hii\ regions (Ferguson {\em et al.\/} 
\cite{FWG96}; Zurita {\em et al.\/} \cite{ZRB00}; Oey {\em et al.\/} 
\cite{Oeyetal07}).  Moreover, across the faces of the galaxies, the \ha\ 
luminosity per unit area of the diffuse ionized gas tracks that of the 
discrete \hii\ regions; that is, it correlates with the Lyman continuum 
production rate per unit area of a galaxy's O~stars (Zurita {\em et al.\/} 
\cite{ZRB00}).  Therefore, O~stars are almost certainly the primary source 
of the WIM's ionization.

\section{Challenges of O~Star Ionization}

The WIM is wide-spread throughout the disk and halo, and its emission line 
spectrum differs significantly from that of the higher density, classical 
\hii\ regions immediately surrounding O~stars.  How can these observed 
properties be reconciled with O~star ionization?

\subsection{Pathways for the Ionizing Photons}

To produce the extensive ionization that is observed, the stellar Lyman 
continuum photons must be able to travel hundreds of parsecs through the 
galactic disk.  This places severe constraints on the distribution and 
column density of the \hi\ within the galaxy.  For example, at a distance 
of 200~pc from an O~star (typical LC luminosity $\approx 10^{49}$ photons 
s$^{-1}$), the ionizing flux is about $2 \times 10^6$ photons s$^{-1}$ 
cm$^{-2}$. A balance of hydrogen ionizations and recombinations will show 
that an ionizing flux at this level will be totally absorbed by a cloud of 
density 1~cm$^{-3}$ or more having a column density of just 
$10^{19}$~cm$^{-2}$.

Models suggest that a fractal \hi\ distribution produced by interstellar 
turbulence (e.g., Elmegreen \cite{e97}; Wood {\em et al.\/} \cite{WHR05}) 
or large cavities (superbubbles/chimneys) created by clustered supernovae 
explosions (e.g., Norman \& Ikeuchi \cite{NI89}) may provide the necessary 
pathways.  For example, 3D radiation transfer models by Wood {\em et 
al.\/} (\cite{wme04}) are illustrated in Figure 1 for ``Str\"omgren 
Spheres'' in density stratified gaseous disks.  For the smooth density 
distribution, the \hii\ region surrounding a single O~star or a small 
cluster of O~stars is confined to the midplane; however, when a portion 
(2/3 in this particular model) of the gas is hierarchically clumped, 
ionizing radiation from a single O~star reaches more than 1~kpc from the 
disk.  Not only models, but direct observations show that superbubbles 
also can be efficient conduits for ionizing photons. One such structure, 
shown in Figure 2, is the Perseus Superbubble, which extends above and 
below Cas OB6 (OCl 352), a cluster of nine O~stars associated with the W4 
\hii\ region in the Perseus spiral arm.  \ha\ emission from the outer 
boundary of this bubble indicates that 40\% or more of the Lyman continuum 
luminosity of the O~star cluster escapes the W4 \hii\ region and ionizes 
gas out to 30$^{\rm o}$ (1200~pc) or more from the stars (Reynolds {\em et 
al.\/} \cite{Reynolds+01}).  However, it is not clear which (if either) of 
these avenues for the transport of ionizing radiation is correct. The 
morphology of the interstellar medium has not yet been characterized, and 
even though some superbubbles are observed to provide extended pathways, 
it is not known whether such cavities are sufficiently prevalent to 
account for the extensive H$^+$.

\begin{figure}[!tb]
\qquad
\includegraphics[width=5cm, angle=0]{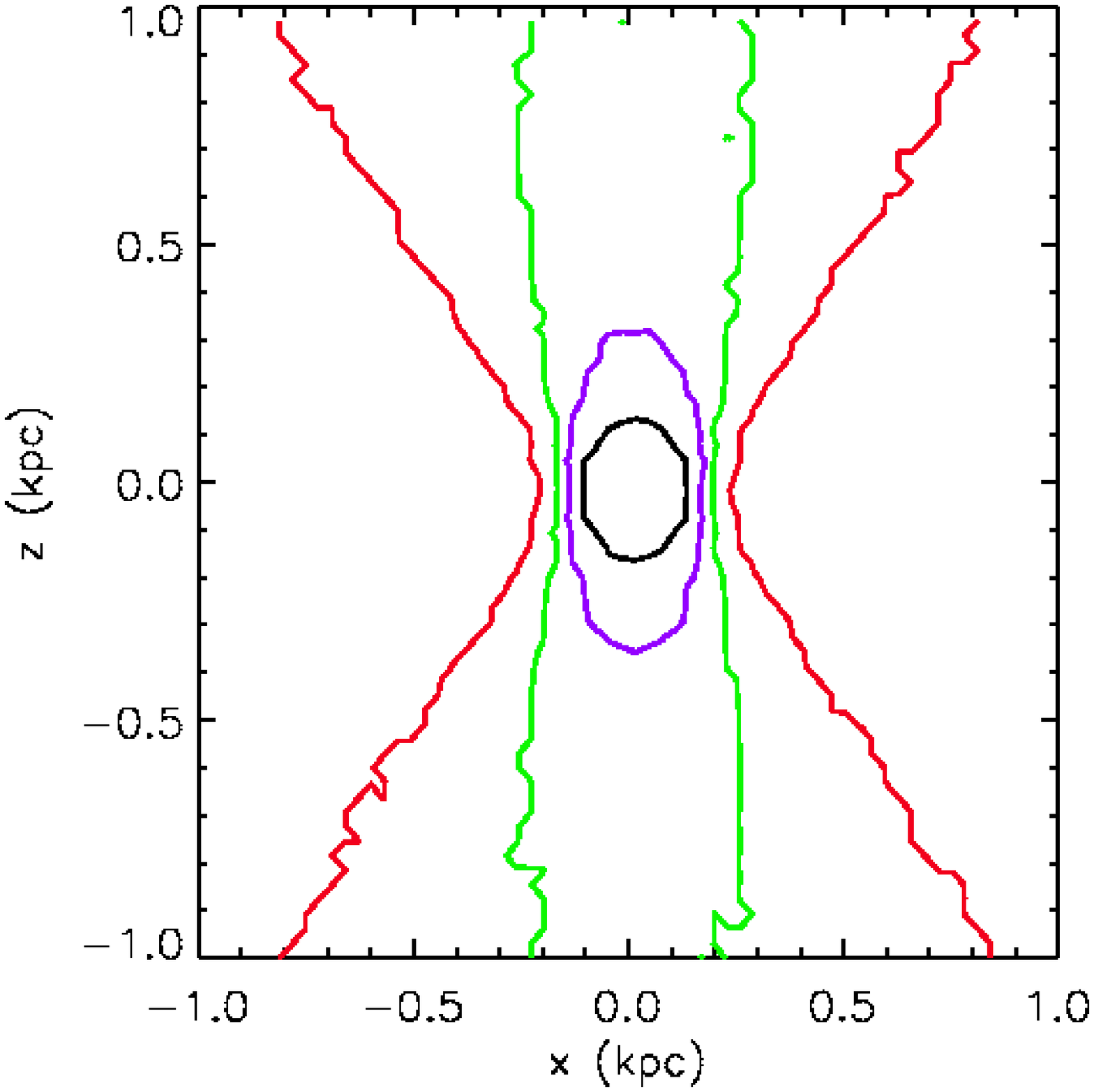}
\includegraphics[width=5cm, angle=0]{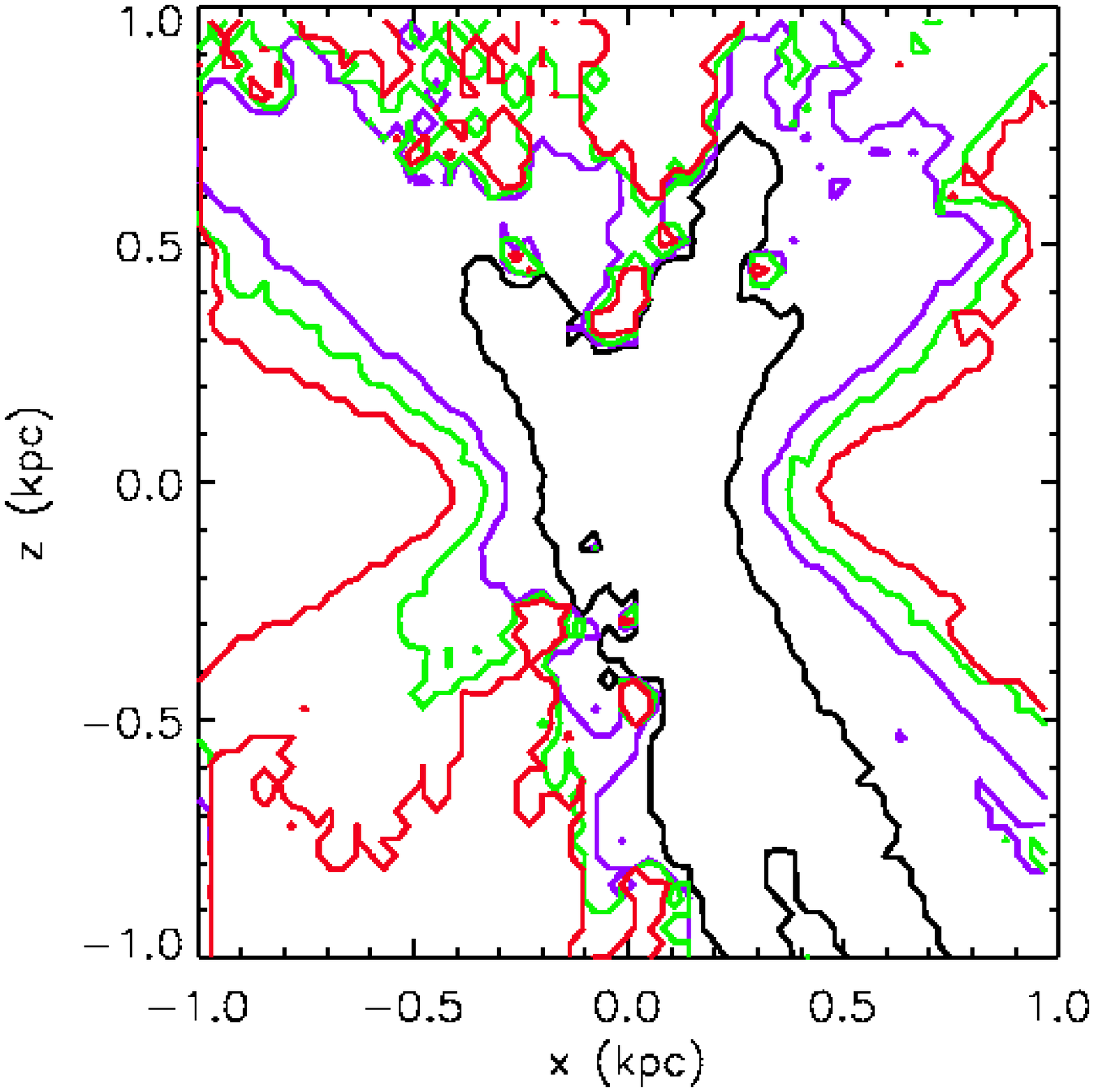}
\caption{Slices through the ionized region for point sources in
smooth (left) and hierarchically clumped (right) density distributions.  
The mean vertical density structure for each simulation is that of a
Dickey-Lockman disk.  From the inner to outer contours the source
luminosities (LC photons s$^{-1}$) are $10^{49}$, $3\times 10^{49}$, 
$5\times 10^{49}$ and $10^{50}$ corresponding to a cluster of 1, 3, 5, and 
10 O~stars, respectively.  A comparison of the contours in the smooth and
clumpy models, shows that the clumped 3D density structure provides
low density paths that allow ionizing photons to reach much larger 
distances from the source.}
\label{fig:wood1}
\end{figure}

\begin{figure}[!tb]
\qquad
\qquad
\quad
\includegraphics[width=8cm, angle=0]{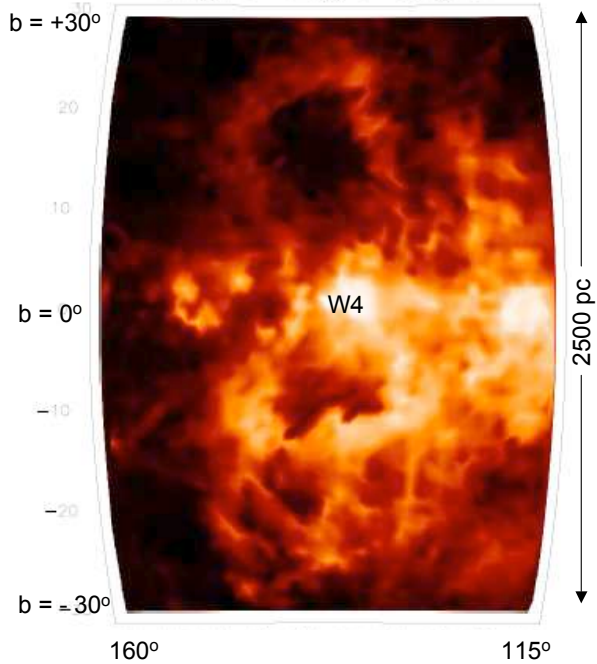}
\caption{A WHAM map of \ha\ emission from the Perseus Superbubble in the 
radial velocity interval $-70$ km s$^{-1} < $V$_{LSR} < -50$ km s$^{-1}$.  
The position of the W4 \hii\ region associated with Cas OB6, the presumed 
source of the ionizing photons, is also indicated.}
\label{perbubble}
\end{figure}

Finally, O~stars alone do not appear to be sufficient to produce the 
diffuse ionization.  A study of edge-on galaxies by Rossa \& Dettmar 
(\cite{RD03}) has shown that the existence of an extended layer of H$^+$ 
in a galaxy appears to require a star formation rate that exceeds a 
specific threshold (corresponding to an FIR surface brightness of $1 - 3 
\times 10^{40}$ ergs s$^{-1}$ per kpc$^2$ of galactic disk).  Above this 
threshold, does the morphology of the interstellar medium change, allowing 
$\approx 1/2$ of the ionizing radiation to escape the star formation 
regions?

\subsection{The Anomalous Emission Line Spectrum}

A second challenge for O~star ionization is the fact that the emission 
line spectrum of the WIM differs significantly from that of the classical 
\hii\ regions that immediately surround O~stars. One of the most 
interesting differences involves the \nii$\lambda$6583/\ha\ and 
\sii$\lambda$6716/\ha\ line intensity ratios.  This is illustrated in 
Figure 3, which compares \sii/\ha\ and \nii/\ha\ in numerous high Galactic 
latitude WIM sightlines with the ratios toward classical \hii\ regions.  
For the \hii\ regions (Fig.~3a), \nii/\ha\ is tightly clustered near 0.25, 
with \sii/\ha\ $\approx 0.1$.  The two exceptions are faint \hii\ regions 
associated with hot, low mass evolved stars (see Madsen {\em et al.\/} 
\cite{MRH06}). On the other hand, for line ratios in directions that 
sample the WIM (Fig.~3b), most of the \nii/\ha\ ratios and nearly all of 
the \sii/\ha\ ratios are significantly larger than those toward the 
classical \hii\ regions.  This result appears to be a characteristic of 
diffuse ionized gas, not just in our Galaxy, but in general (see, e.g., 
Rand \cite{Rand98}; T{\" u}llmann \& Dettmar \cite{TD00}; Hoopes \& 
Walterbos \cite{HW03}).  Discussions of other spectral differences can be 
found in T{\" u}llmann \& Dettmar \cite{TD00}, Madsen {\em et al.\/} 
\cite{MRH06}, and Haffner {\em et al.\/} \cite{haffner09}.

There is strong evidence that these enhanced forbidden line intensities 
(relative to \ha) are due primarily to higher temperatures in the WIM 
compared to the classical \hii\ regions.  The \nii$\lambda$6583/\ha\ 
intensity ratio is given by

\begin{equation} 
\frac{[\rm{N~\textsc{II}}]}{\rm{H}\alpha} =
1.62\times10^5~T_4^{0.4}~e^{-2.18/T_4}
\left(\frac{\rm{N}^+}{\rm{N}}\right) \left(\frac{\rm{N}}{\rm{H}}\right)  
\left(\frac{\rm{H}^+}{\rm{H}}\right)^{-1}, 
\label{ch1eq:niieq}    
\end{equation}

\begin{figure}[!tb]
\qquad
\includegraphics[width=6.5cm, angle=270]{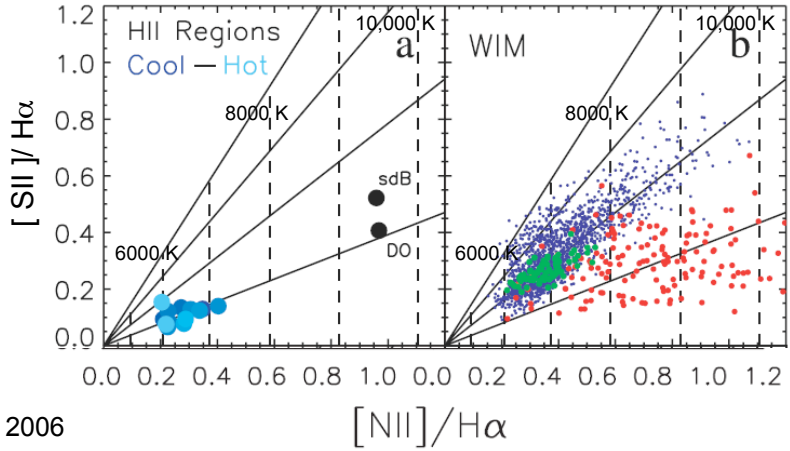}
\caption{WHAM observations of \nii/\ha\ versus \sii/\ha\ toward classical
OB star \hii\ regions (left) and in directions that sample the WIM
(right). The dashed vertical lines represent lines of constant temperature
(from eq. 4.1) with 5000~K $< T <$\ 10~000~K. The four sloped solid
lines represent, with increasing slope, values of constant S$^+$/S $ =$ 
0.25, 0.50, 0.75, and 1.0 (eq. 4.2). For a more detailed description, see 
Madsen {\em et al.\/} \cite{MRH06}.}
\label{fig:siivsnii}
\end{figure}

\noindent where $T_4$ is the electron temperature $T$ in units of 
$10^4$~K, N/H is the gas phase abundance of nitrogen relative to hydrogen, 
and N$^+$/N and H$^+$/H are the ionization fractions of N and H, 
respectively.  Because H$^+$/H is nearly equal to N$^+$/N (both are near 
unity) in the ionized regions, local variations in \nii/\ha\ must be due 
essentially to variations in temperature (see discussion in Madsen {\em et 
al.\/} \cite{MRH06}).  This conclusion has been confirmed by observations 
of other emission line ratios, in particular, \oii/\ha\ (Otte {\em et 
al.\/} \cite{OGR02}; Mierkiewicz {\em et al.\/} \cite{Mierkiewicz+06}) and 
\nii$\lambda$5755/\nii$\lambda$6583 (Madsen {\em et al.\/} \cite{MRH06}).

Because the excitation energies for the \sii$\lambda6716$ and 
\nii$\lambda6583$ emission lines are nearly identical, the expression for 
\sii/\ha\ has nearly the same dependence on $T$ as the expression for 
\nii/\ha, so that

\begin{equation}
\frac{[\rm{S~\textsc{II}}]}{[\rm{N~\textsc{II}}]} = 4.62~e^{0.04/T_4}~    
        \left(\frac{\rm{S}^+}{\rm{S}}\right)
        \left(\frac{\rm{S}}{\rm{H}}\right)
    \left[\left(\frac{\rm{N}^+}{\rm{N}}\right)
    \left(\frac{\rm{N}}{\rm{H}}\right)\right]^{-1}.
\label{ch1eq:siieq}
\end{equation}

\noindent However, due to its relatively low ionization potential (23.4~eV 
for S$^+$ compared to 29.6~eV for N$^+$), S$^+$ is not necessarily the 
dominant ion within the ionized regions.  As a result, observed variations 
in \sii/\nii\ are primarily a measure of variations of S$^+$/S.

From the above discussion, the data in Figure 3 can be interpreted as 
measurements of $T$ and S$^+$/S within the ionized regions.  The data
show that not only is the WIM generally warmer than the \hii\ regions, but 
that within the WIM there are substantial variations in both the 
temperature and ionization state from one line of sight to the next.

\section{Why is the WIM Warmer than the Classical \hii\ Regions?}

The variations in S$^+$/S could be a measure of differences in the 
ionization parameter, that is, the S$^+$-ionizing photon density to 
electron density ratio, within the ionized regions (Mathis \cite{m86}).  
The significant variations in temperature are more difficult to 
understand.  Photoionization models suggest that some temperature 
increases could result from the hardening of the ionizing spectrum as the 
radiation passes through density bounded \hii\ regions before reaching the 
more distant WIM (Wood \& Mathis \cite{WM04}; Wood {\em et al.\/} 
\cite{WHR05}).  However, this mechanism does not explain the highest 
values of \nii/\ha, i.e., the large range in temperatures that is 
observed.  To produce the higher temperatures, an additional, non-ionizing 
source of heat must be added (Wood \& Mathis \cite{WM04}).

One clue about the nature of this supplemental heating is the observed 
strong anticorrelation between \nii/\ha\ (temperature) and the \ha\ 
intensity (gas density).  This anticorrelation is apparent in observations 
of our Galaxy (e.g., Madsen {\em et al.\/} \cite{MRH06}) and other 
galaxies, not just with increasing distance $|z|$ from the midplane, but 
also in observations at constant $|z|$ (e.g., Rand \cite{Rand98}).  This 
behavior could be explained if there were a heat source with a heating 
rate per unit volume proportional to the first power of the density, or 
did not depend upon density at all.  Then, at sufficiently low densities, 
such a source would dominate over photoionization heating (which is 
proportional to $n^2$), producing the observed inverse relationship 
between temperature and density (see Reynolds {\em et al.\/} 
\cite{RHT99}).  The required heating rate is $\sim 10^{-26} - 10^{-25}$ 
erg s$^{-1}$ per H$^+$, or $\sim 10^{-27}$ erg s$^{-1}$ cm$^{-3}$ (Wood 
\& Mathis \cite{WM04}; Reynolds {\em et al.\/} \cite{RHT99}).  Possible 
sources include photoelectric heating by dust (Weingartner \& Draine 
\cite{WD01}), dissipation of turbulence (Minter \& Spangler \cite{MS97}), 
and magnetic reconnection (Raymond \cite{raymond92}).

\section{Conclusions}

The thick layers of warm, low density H$^+$ in spiral galaxies contain 
clues about the structure of the interstellar medium and the transport of 
energy away from regions of star formation at the midplane.  Some progress 
has been made in understanding this gas, but many fundamental questions 
remain:  How does the ionizing radiation propagate through the disk? Why 
is the H$^+$ layer so thick?  What is the distribution of the WIM within 
the interstellar medium? What is its relationship to the other phases of 
the medium? What is the source of the elevated temperatures in the WIM?  
Comparisons of the observed distribution, motions, and line ratios of the 
WIM with predictions of models have begun to address some of these 
questions (e.g., Hill {\em et al.\/} \cite{Hill08}; Wood \& Mathis 
\cite{WM04}). Perhaps a next step is to compare the observations to more 
sophisticated multi-phase, dynamical models that included sources of 
heating and ionization with 3D radiation transfer.  The observations would 
constrain model parameters, and the models could predict properties of the 
gas to guide future observations.\\

\thanks{RJR, LMH and AH acknowledge support from the National Science 
Foundation (NSF) through AST06-07512.  GJM is supported by the University 
of Sydney Postdoctoral Fellowship Program and the NSF grant AST04-01416.}

\end{document}